\documentstyle[prl,aps,float,multicol,epsf]{revtex}

\floatstyle{plain}  

\newfloat{FIG4}{t}{lop}

\begin{document}

\draft
\title{ Detection of Coulomb Charging around an Antidot in the Quantum Hall
Regime }

\author{M.~Kataoka,$^{1}$ C.~J.~B.~Ford,$^{1}$ G.~Faini,$^{2}$
D.~Mailly,$^{2}$ M.~Y.~Simmons,$^{1}$ D.~R.~Mace,$^{1,\ast}$ C.-T.~Liang,$^{1}$
and D.~A.~Ritchie$^{1}$}

\address{
$^{1}$Cavendish Laboratory, Madingley Road, Cambridge CB3 0HE, United Kingdom
}

\address{
$^{2}$Laboratoire de Microstructures et de Microelectronique - CNRS,
196,~Avenue Henri Ravera, 92220~Bagneux, France }

\date{\today}

\maketitle

\widetext
\begin{abstract}
\leftskip 54.8pt
\rightskip 54.8pt

We have detected oscillations of the charge around a potential hill (antidot)
in a two-dimensional electron gas as a function of a large magnetic field
$B$. The field confines electrons around the antidot in closed orbits, the
areas of which are quantised through the Aharonov-Bohm effect. Increasing $B$
reduces each state's area, pushing electrons closer to the centre, until
enough charge builds up for an electron to tunnel out. This is a new form of
the Coulomb blockade seen in electrostatically confined dots. Addition and
excitation spectra in DC bias confirm the Coulomb blockade of tunneling.


\end{abstract}

\begin{multicols}{2}
\narrowtext

This paper addresses the fundamental question of whether charging can occur
in an open system. Coulomb blockade (CB) of tunnelling is generally only
observed in electrostatically confined ``dots'' where there is only partial
transmission through the entrance and exit constrictions. It has recently
been seen when one constriction is open \cite{openCB}, when both
constrictions transmit exactly one one-dimensional (1D) channel
\cite{Marcus}, or when some transmitted channels are decoupled from trapped
states \cite{CTLopenCB}. However, an unambiguous demonstration requires a
completely open system, such as an antidot, which is a potential hill in a
two-dimensional electron gas (2DEG). When a magnetic field $B$ is applied
perpendicular to the 2DEG, a set of states, discrete in position and energy,
is formed around the antidot, for each Landau level (LL). Aharonov-Bohm (AB)
conductance oscillations arising from resonances through such states have
been studied extensively in the integer and fractional quantum Hall (QH)
regimes \cite{Chris,Sachrajda1,Goldman,Mace,Franklin,Maasilta}.  It has often
been assumed that CB does not occur with antidot states because, as charge
tries to build up, the system must immediately respond to screen it. However,
pairs of AB oscillations from the two spins of the lowest LL were found to
lock in antiphase, and this was attributed to charging
\cite{Chris,Sachrajda1}. In a dot system, it was suggested that the charging
of edge channels is responsible for a similar regularity of the
magnetoconductance peaks \cite{Dharma,Sachrajda2}.

\begin{figure}        

\epsfxsize=85truemm		  

\centerline{\epsffile{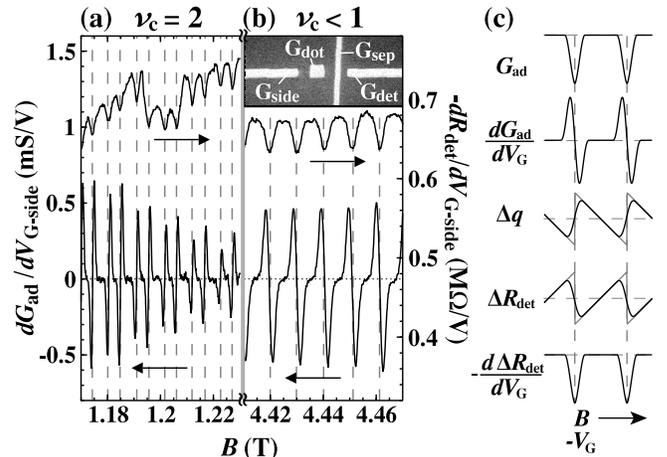}}       


\caption{ 
$dG_{\rm{ad}}/dV_{\rm{G-side}}$ of the antidot circuit and
$-dR_{\rm{det}}/dV_{\rm{G-side}}$ of the detector circuit with the gate
voltage on G$_{\rm{side}}$ modulated in two different regimes: (a) $\nu
_{\rm{c}} = 2$ and (b) $\nu _{\rm{c}} < 1$. Vertical dashed lines show the
alignment of the dips in the detector signal with zeros in the
transconductance oscillations. Inset: SEM micrograph of a device prior to
second metallisation. (c) Illustration of the relation between various
lineshapes. Grey lines in $\Delta q$ and $\Delta R_{\rm{det}}$ are the ideal
case, and black curves represent thermally broadened lineshapes.
 }       

\label{figDet}         
\end{figure}

The aim of the present work was to detect such charge oscillations of an
antidot, utilising a non-invasive voltage probe similar to that employed by
Field {\em et al.\/} \cite{Field}. They fabricated a 1D 
constriction as a charge detector next to a dot but in a different circuit
separated from it by a narrow gate.  When the constriction was nearly pinched
off, its resistance was very sensitive to potential variations nearby, and
hence it could detect charge oscillations in the dot.  We have fabricated a
similar device with an antidot instead of a dot (see inset to
Fig.~\ref{figDet}(b)). A charging signal with the same period as the AB
oscillations in the conductance $G_{\rm ad}$ is clearly visible. The lineshape
and phase show that CB of tunnelling through the antidot is occurring.
DC-bias measurements are used to measure addition and excitation spectra,
confirming this interpretation. The charging energy saturates at high $B$ and
the single-particle (SP) energy spacing varies as $1/B$.


The devices were fabricated from a GaAs/AlGaAs heterostructure with a 2DEG of
sheet carrier density $2.2 \times 10^{15}$~m$^{-2}$ and mobility
370~m$^{2}$/Vs after illumination by a red LED. An SEM micrograph of a device
is shown in the inset to Fig.~\ref{figDet}(b). A square dot gate
(G$_{\rm{dot}}$), 0.3~$\mu$m on a side, was contacted by a second metal layer
evaporated on top of an insulator (not shown), to allow independent control
of gate voltages. The lithographic widths of the antidot and detector
constrictions were 0.45~$\mu$m and 0.3~$\mu$m respectively. All constrictions
showed good 1D ballistic quantisation at $B = 0$. A voltage of $-4.5$~V on
the separation gate (G$_{\rm{sep}}$), of width 0.1~$\mu$m, divided the 2DEG
into separate antidot and detector circuits. The detector gate
(G$_{\rm{det}}$) squeezed the detector constriction to a resistance between
0.1 and 5~M$\Omega$ to make it very sensitive to nearby charge. To maximise
the sensitivity transresistance measurements were made by modulating the
dot-gate voltage (or the voltage on the side-gate G$_{\rm{side}}$) at 10~Hz
with 0.5~mV rms and applying a DC current of 1~nA through the detector
constriction. Simultaneously, the transconductance of the antidot circuit was
measured with a 10~$\mu$V DC source-drain bias, when necessary. The
experiments were performed in a dilution refrigerator with a base temperature
below 50~mK.

Figure~\ref{figDet} shows the transresistance
$-dR_{\rm{det}}/dV_{\rm{G-side}}$ (transconductance
$dG_{\rm{ad}}/dV_{\rm{G-side}}$) vs $B$ of the detector (antidot) circuit in
two different field regions: (a) $\nu _{\rm{c}} = 2$ and (b) $\nu _{\rm{c}} <
1$, where $\nu _{\rm{c}}$ is the filling factor in both antidot
constrictions, which were determined from $G_{\rm{ad}}$. The filling factors
in the bulk 2DEG were $\nu_{\rm{b}}=7$ and 2, respectively. The oscillations
in $G_{\rm ad}$ occur as SP states around the antidot rise up through the
Fermi energy $E_{\rm F}$. The AB effect causes the overall period $\Delta B$
to be $h/eS$, where $S$ is the area enclosed by the state at $E_{\rm F}$. 
The curve in (a) has pairs of spin-split peaks, whereas in (b) only one spin
of the lowest LL is present. The dips in $-dR_{\rm{det}}/dV_{\rm{G-side}}$
correspond to a saw-tooth in the change $\Delta R_{\rm{det}}$ from the
background resistance (see Fig.~\ref{figDet}(c)). Thus the net charge $\Delta
q$ nearby suddenly becomes more positive (making the effective gate voltage
less negative) whenever the antidot comes on to resonance (since the dips
line up with the zeros in $dG_{\rm{ad}}/dV_{\rm{G-side}}$).  Hence we
conclude that this charge oscillation is associated with states near the
antidot. A second sample showed very similar results.  

We explain the charging as follows.  As $B$ increases, all the states
encircling the antidot move inwards, reducing their areas to keep the
flux enclosed constant, and hence a net charge $\Delta q$ builds up in the
region around the antidot. This resembles CB in a dot \cite{Beenakker}. At
low bias, the electron in the highest occupied state cannot escape until
$\Delta q$ reaches $-e/2$, then it tunnels out to a nearby lead or into a
localised state in the bulk, and $\Delta q$ becomes $+e/2$. At this point
charge can move easily through the antidot, and so each dip in the detector
signal lines up with such a conductance resonance, as found experimentally
(Fig.~\ref{figDet}). There is no electrostatically confined region around the
antidot, so charging seems impossible \cite{Goldman}. However, electrons are
magnetically confined to the antidot and the rigidity of the
quantum-mechanical orbitals prevents charge relaxation. Other states further
away from the antidot might try to screen the charge build-up. However, those
in the same LL have a fixed density once it is full, and so cannot screen. 
Also, due to the discreteness of the SP states, rearrangement of charge below
$E_{\rm F}$ within the partially filled region near the antidot can only
cause discrete changes in the charge, and would probably cost too much
interaction energy. One might speculate that the detector would pick up not
the charging of the antidot but the change in screening by SP states near
$E_{\rm{F}}$ because they could adjust their areas or the wavefunction could
leak out to the other edges on resonance \cite{Field2}. However, such
screening should be symmetric around the resonances. Therefore the
transresistance would be the derivative of periodic dips or peaks, not of a
saw-tooth as seen in our measurements.


\begin{figure}    

\epsfxsize=85truemm  

\centerline{\epsffile{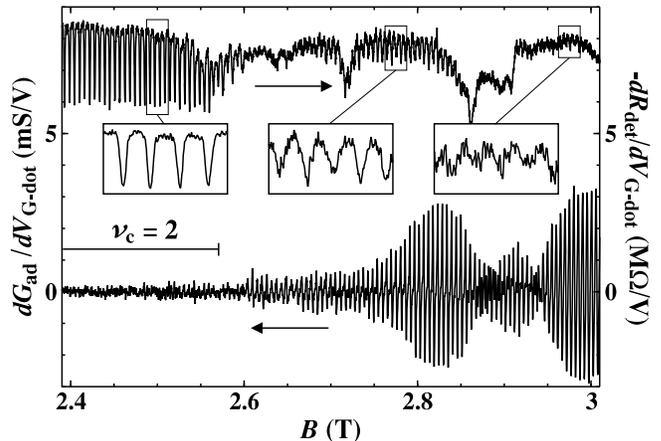}}  


\caption { 
Antidot and detector signals with the antidot voltage modulated, in the
regime of pure $h/2e$ AB oscillations. The amplitude of the detector
oscillations (upper curve) suddenly decreases at the onset of the
oscillations in the antidot circuit (lower curve). 
 }   

\label{fig3T}      
\end{figure}

The charging of the antidot is not dependent on the presence of conductance
oscillations in the antidot circuit. Thus it is still possible to observe the
signal with no applied bias in the antidot circuit, or when the side-gate
voltage is zero so that there is no tunnelling between that edge and the
antidot. Indeed, as shown in Fig.~\ref{fig3T}, the dips in the detector
signal become large and sharp when the antidot constrictions are set to a QH
plateau ($\nu _{\rm{c}}=2$ in this case), where the antidot states are
decoupled from the extended edge states. Away from the QH plateau, since the
states are coupled to the current leads, electrons' wavefunctions penetrate
into the leads, reducing the effective maximum charge on the antidot and
leading to weaker charging, i.e. smaller charging energy.

\begin{figure}     

\epsfxsize=85truemm 

\centerline{\epsffile{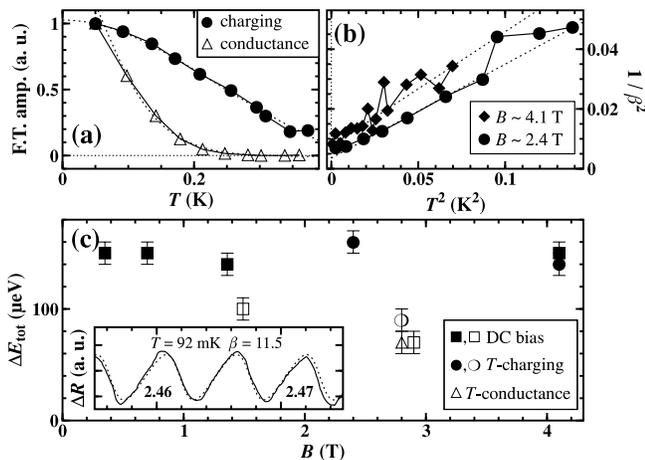}}  


\caption {
(a) Temperature dependence of the Fourier amplitude of $h/2e$ oscillations in
conductance (not transconductance) (triangles, $B \approx 2.8$~T) and
charging (circles, $B \approx 2.5$~T). The amplitudes were calculated by
taking the square root of the Fourier power spectrum integrated around the
$h/2e$ peak. Dashed lines are fits to thermally broadened Fermi-liquid
theory. (b) $\beta$ (see text) $vs$ $T^{2}$ for $h/2e$ (circles) and $h/e$
(diamonds) AB oscillations. The straight lines are fits. (c) $\Delta E_{\rm
tot}$ obtained from the measurements shown. Filled (open) symbols correspond
to $G_{\rm ad}$ on (below) the $\nu _{\rm c}=2$ plateau, except for the
highest $B$ data which is on the $\nu _{\rm c}=1$ plateau. Inset: integral of
the detector oscillations. The dashed line is a fit as described in the text.
 }        

\label{figTdp}           

\end{figure}

Around $B=3$~T the spin-splitting of the peaks becomes exactly half the
period, and the amplitudes of the two peaks in each pair become identical,
giving what appear to be $h/2e$ AB oscillations (see Fig.~\ref{fig3T})
\cite{Chris,Sachrajda1}.  We have investigated the temperature dependence of
both the charging and conductance signals in this regime. The Fourier
transforms of the charging signal appearing at around 2.5~T in
Fig.~\ref{fig3T} and the $G_{\rm{ad}}$ oscillations at around 2.8~T, measured
separately, decrease at different rates (see Fig.~\ref{figTdp}(a)). Thermally
broadened Fermi-liquid theory for sinusoidal oscillations \cite{Franklin}
gives a good fit for $G_{\rm ad}$ with an energy level spacing of $70~\mu$eV.
The conductance oscillations are suppressed at high temperature because of
thermal broadening of the edge channels around the side gates at $E_{\rm{F}}$
when the thermal energy becomes comparable to the sum of the SP energy
spacing and the charging energy $e^2/C$ (if CB occurs), where $C$ is the
total capacitance of the antidot. For the charging signal, since the
oscillations are not sinusoidal, a more detailed model \cite{Beenakker} is
required than that used above. Here, we assume that the detector is only
sensitive to thermal excitation which adds or removes electrons around the
antidot, but not to excitation between SP states.


The electrochemical potential of the antidot $\mu_{\rm{ad}}(N,B)$ is the
energy required to add an electron to the lowest unoccupied state, which
encloses, say, $N$ flux quanta $h/e$. Then the probability that thermal
excitation moves an electron from a lead at chemical potential $\mu$ to that
state is given by the Fermi function $f \left( \mu_{\rm{ad}}(N,B) - \mu
\right) $.  For one period $- \frac{\Delta B}{2} < B < \frac{\Delta B}{2}$,
where the centre of the charge transition is at $B=0$, the blurred saw-tooth
charge oscillation can be written as $\Delta q (B) = -e\left(B/ \Delta B
+f\left(\mu_{\rm{ad}} \left( N,B \right) - \mu \right) -\frac{1}{2}\right)$.
Since the charging energy is parabolic in the net charge, and hence varies as
$(B \pm \frac{\Delta B}{2})^2$ depending on which state is occupied, it can
be shown simply that $\mu_{\rm{ad}}(N,B) - \mu = \Delta E_{\rm tot} B /
\Delta B$, where  $\Delta E_{\rm tot} = \Delta E + e^{2}/C$. Here, $\Delta E$
is the average energy spacing of adjacent states (of whichever spin), equal
to $\Delta E_{\rm{sp}}/2$ when both spins encircle the antidot; $\Delta
E_{\rm{sp}}$ is the energy spacing of adjacent SP states of the same spin. 
For $h/2e$ oscillations, we assume that a spin-down state lies midway in
energy between spin-up states. For $\nu_{\rm c} \leq 1$, $\Delta E = \Delta
E_{\rm{sp}}$. Thus 
 $f \left( \mu_{\rm{ad}}(N,B) - \mu \right) = \left(1+\exp \left( -\beta B/
\Delta B \right) \right)^{-1}$
 where $\beta = \Delta E_{\rm tot} / k_{\rm{B}}T^{\ast}$. Here
$T^{\ast}=\sqrt{T^{2}+\Gamma ^{2}}$ is the effective temperature, to account
for an intrinsic broadening $\Gamma$ at low temperatures due to the AC
excitation voltage and the finite lifetime of the states. The integral of the
detector signal with respect to $B$ (approximately equivalent to the
integral with respect to $-V_{\rm G-dot}$) was fitted to $\Delta q (B)$,
after subtracting the background slope (inset to Fig.~\ref{figTdp}(c)). From
the fit at various temperatures (Fig.~\ref{figTdp}(b), circles), we obtained
$\Delta E_{\rm tot} = 160~\mu$eV. We could not measure the temperature
dependence in the region $B \approx 2.8$~T in Fig.~\ref{fig3T} due to the
small charging signal. However, a fit to the data at $T \approx 50$~mK gave
$\Delta E_{\rm tot} = 90~\mu$eV, assuming that $\Gamma$ does not change. The
temperature dependence of $h/e$ oscillations where $\nu _{\rm{c}}$ was just
less than one ($B \approx 4.1$~T, diamonds in Fig.~\ref{figTdp}(b)) gave
$\Delta E_{\rm tot} = 140~\mu$eV. These energies are plotted in
Fig.~\ref{figTdp}(c) and will be discussed below.

A further way of measuring the energy spacing is to apply a DC bias 
\cite{Foxman}. Fig.~\ref{figDC} shows greyscale plots of the DC-bias
dependence of AB oscillations in the differential conductance (measured with
a 5~$\mu$V rms AC (10~Hz) source-drain voltage in addition to the DC bias),
for the values of $\nu _{\rm{c}}$ shown. In (a) and (b), peaks are shown in
black, since resonant transmission occurs due to inter-LL scattering
\cite{Mace}.  This is not present at higher $B$; instead, resonant
backscattering gives dips (shown in black in (c) and (d)).  (a)--(c) show
sets of spin-split resonances. In (a), where spin splitting is poorly
resolved, adjacent peaks cross at $250~\mu$V or $50~\mu$V. As energies,
since this is an addition spectrum, these correspond to $e^2/C + \Delta
E_{\rm sp} - E_{\rm Z}$ and $e^2/C + E_{\rm{Z}}$ respectively, where $E_{\rm
Z}$ is the Zeeman splitting. Thus the {\em average} energy is just $\Delta
E_{\rm{tot}}$. This enables a comparison of energies at various $B$ (see
Fig.~\ref{figTdp}(c)).  At higher $B$, spin-splitting becomes obvious
(Figs.~\ref{figDC}(b) and (c)), but the  crossings give similar $\Delta
E_{\rm{tot}}$.

The DC bias at which states of different spin cross gives an upper limit for
$e^2/C$, and this limit increases with $B$, as does $E_{\rm Z}$.  It is
likely that the charging energy is small at low $B$, since the magnetic
confinement is weak; indeed, the charging signal is hard to see for
$B<0.6$~T.  However, at $B=1.4$~T, in the middle of Fig.~\ref{figDC}(c),
$\Delta E_{\rm tot}$ drops rapidly by 30\% (open symbol in
Fig.~\ref{figTdp}(c)).  This corresponds to the field at which the
conductance falls off the $\nu _{\rm c}=2$ plateau (for these particular gate
voltages). The figure shows a similar drop (for the gate voltages used in the
temperature dependence described above), around 2.6~T, also corresponding to
moving off the $\nu _{\rm c}=2$ plateau. Temperature dependences of the
conductance and charging oscillations there confirm the DC bias result. 
There is no reason why $\Delta E_{\rm{sp}}$ should change so suddenly.  These
drops occur when the coupling of the antidot to the leads increases,
reducing the charging energy, as described above.

In Fig.~\ref{figDC}(c) additional parallel lines appear around the smaller
diamonds, offset by $60~\mu$V in DC bias. We interpret these as arising from
tunnelling via the first excited state of the antidot, which is $\Delta
E_{\rm sp}-E_{\rm Z}$ higher in energy. Similar lines are not resolved around
the larger diamonds since the spacing is just $E_{\rm{Z}}$. The observation
of this excitation spectrum confirms that there is a Coulomb blockade of
tunnelling through the antidot.

For a constant potential slope, $\Delta E_{\rm{sp}}$ should vary as $1/B$. 
At $B=0.35$~T, $\Delta E_{\rm{tot}}=150~\mu$eV and $e^2/C<50~\mu$eV (the
upper limit from the DC-bias measurements), so $200~\mu$eV$<\Delta
E_{\rm{sp}}<300~\mu$eV. Thus at $B=1.4$~T we expect $50~\mu$eV$<\Delta
E_{\rm{sp}} < 80~\mu$eV. This is close to the value $\Delta E_{\rm sp}
\approx 100~\mu$eV obtained from the addition and excitation spectra at
1.4~T, which also give $E_{\rm{Z}} \approx 35~\mu$eV, in good agreement with
$g\mu _{\rm{B}}B$ with $g=0.44$ for electrons in GaAs. From
Fig.~\ref{figDC}(c), $e^{2}/C=\Delta E_{\rm{tot}}-\Delta E_{\rm{sp}}/2$ falls
from $\approx 100~\mu$eV on the plateau to $\approx 65~\mu$eV when the
antidot is coupled to the leads.  When on the plateau, $e^2/C$ appears to
saturate at $\approx 150~\mu$eV above $B\approx 2$~T, since by then the
states around the antidot are well defined and so the full $\pm e/2$ charge
can build up, with the capacitance fairly constant.  Maasilta and Goldman
\cite{Maasilta} found from the lineshapes of individual peaks at $\nu=1$ and
$\frac{1}{3}$ that $\Delta E_{\rm{tot}}$ was almost constant, but interpreted
this as a self-consistent variation of the potential slope, with no CB.  In
our picture, the constancy of $\Delta E_{\rm tot}$ comes from the interplay
of $e^2/C$ and $\Delta E_{\rm{sp}}$.

In summary, we have fabricated a charge detector in close proximity to
an antidot. The antidot is seen to discharge each time a state around the
antidot comes on to resonance, showing that there is a Coulomb blockade of
tunnelling via the antidot. We have measured addition and excitation spectra,
confirming this interpretation. The charging energy drops whenever there is
coupling to the leads, as the charge becomes delocalised. This is the first
conclusive demonstration of charging in an open system. It arises from the
rigidity of the quantum-mechanical wavefunction, as for an electron in an
atom. It must form part of the explanation for the pure $h/2e$ AB
oscillations \cite{Chris}.

This work was funded by the UK EPSRC. We thank C.~H.~W.~Barnes and
C.~G.~Smith for useful discussions. M.~K. acknowledges financial support from
Cambridge Overseas Trust.

$^{\ast}$Present address: The Technology Partnership PLC, Melbourn Science Park,
Melbourn, SG8 6EE, UK.

\end{multicols}

\widetext

\setlength{\textfloatsep}{1mm}

\begin{FIG4}


\begin{figure}[!t]       

\epsfxsize=178truemm   

\centerline{\epsffile{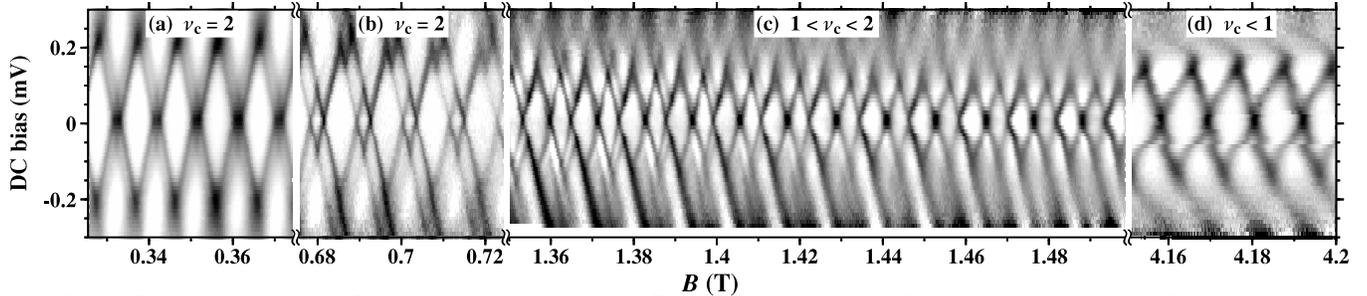}}

\caption {   
Greyscale plots of DC-bias dependence of the differential conductance of the
antidot at various $B$. The same gate voltages were used throughout. Dark
regions correspond to positions of peaks or dips. The background variation in
the signal was subtracted to increase the contrast.
 }     

\label{figDC}   

\end{figure} 

\end{FIG4}


\begin{references}

\centerline{\bf References}

\bibitem{openCB}        C.~Pasquier {\em et al.\/}, Phys. Rev. Lett. {\bf
70}, 69 (1993); C.~H.~Crouch {\em et al.\/}, Superlatt. and Microstr. {\bf
20}, 377 (1996).

\bibitem{Marcus}        S.~M.~Cronenwett {\em et al.\/}, Phys. Rev. Lett.
{\bf 81}, 5904 (1998).

\bibitem{CTLopenCB}     C.~T.~ Liang {\em et al.\/}, Phys. Rev. Lett. {\bf
81}, 3507 (1998).

\bibitem{Chris}		C.~J.~B.~Ford {\em et al.\/}, Phys. Rev. B {\bf 49},
17456 (1994).

\bibitem{Sachrajda1}	A.~S.~Sachrajda {\em et al.\/}, Phys. Rev. B {\bf
50}, 10856 (1994).

\bibitem{Goldman}	V.~J.~Goldman and B.~Su, Science {\bf 267}, 1010
(1995).

\bibitem{Mace}		D.~R.~Mace {\em et al.\/}, Phys. Rev. B {\bf 52},
R8672 (1995).

\bibitem{Franklin}	J.~D.~F.~Franklin {\em et al.\/}, Surf. Sci. {\bf
361/362}, 17 (1996).

\bibitem{Maasilta}	I.~J.~Maasilta and V.~J.~Goldman, Phys. Rev. B {\bf
57}, R4273 (1998).

\bibitem{Dharma}	M.~W.~C.~Dharma-Wardana, R.~P.~Taylor, and
A.~S.~Sachrajda, Solid State Commun. {\bf 84}, 631 (1992).

\bibitem{Sachrajda2}	A.~Sachrajda {\em et al.\/}, Surf. Sci. {\bf305}, 527
(1994).

\bibitem{Field}		M.~Field {\em et al.\/}, Phys. Rev. Lett. {\bf 70},
1311 (1993).

\bibitem{Beenakker}     C.~W.~J.~Beenakker, Phys. Rev. B {\bf 44}, 1646
(1991).

\bibitem{Field2}	M.~Field {\em et al.\/}, Phys. Rev. Lett. {\bf 77},
350 (1996).


\bibitem{Foxman}	E.~B.~Foxman {\em et al.\/}, Phys. Rev. B {\bf 47},
10020 (1993).


\end{references}
\end{document}